\documentclass{article}

\usepackage[utf8]{inputenc}
\usepackage{amsmath,amssymb, xcolor, subcaption,hyperref, amsfonts, mathtools}
\usepackage{authblk}
\usepackage[scale=0.8]{geometry} 

\usepackage{authblk}

\definecolor{blue1}{RGB}{0,76,153}
\definecolor{green1}{RGB}{1,109,57}
\definecolor{red1}{RGB}{153,0,0}

\title{Enforcing Thermodynamic Consistency in Parameter Optimization and Bayesian Inference of Biochemical Reaction Networks}
\author[1,+]{Federica Milinanni\thanks{Corresponding author: \texttt{federica.milinanni@northwestern.edu}}}
\author[2,+]{Andrei Kramer-Miehe}
\author[3]{Jan Glaubitz}
\author[2]{Olivia Eriksson}
\affil[1]{Department of Industrial Engineering and Management Science, Northwestern University}
\affil[2]{School of Electrical Engineering and Computer Science, Science for Life Laboratory, KTH Royal Institute of Technology}
\affil[3]{Department of Mathematics, Linköping University}

\affil[+] {Have contributed equally to this work as first authors.}

\usepackage{natbib}
\usepackage{graphicx}

\def\bfx{\mathbf{x}}
\def\bfy{\mathbf{y}}
\def\bfe{\mathbf{e}}
\def\bfu{\mathbf{u}}

\def\bft{\boldsymbol{\theta}}
\def\bfg{\mathbf{g}}

\begin{document}

\maketitle
\begin{abstract}
    Subcellular signaling pathways regulating, e.g., cell division or synaptic plasticity, are often modeled as dynamical networks of reactions satisfying the law of mass action. To remain physically grounded, these models must satisfy different physical conservation laws, such as conservation of energy. This gives rise to thermodynamic constraints formalized by the Wegscheider conditions, which ensure that the system is in detailed balance at thermodynamic equilibrium.

    Explicitly imposing Wegscheider conditions requires identifying cycles in the reaction network. This can become computationally cumbersome, especially in large, complex systems. Furthermore, even when these constraints are explicitly available, enforcing them requires choosing a reparameterization of the network, which may introduce asymmetries among the parameters and affect numerical conditioning. Moreover, this approach often causes inverse uncertainty quantification methods like Markov chain Monte Carlo (MCMC) to struggle.

    Instead of imposing Wegscheider conditions directly in the parameter space, we propose an approach for promoting physical consistency by penalizing parameters that violate detailed balance.

    We first illustrate the approach on a simple reaction network containing a single cycle. We show that parameter values inferred without thermodynamic constraints can reproduce the experimental data seemingly accurately, while substantially violating the Wegscheider condition. In contrast, the proposed approach preserves the agreement with the observed dynamics while yielding posterior samples that approximately satisfy the thermodynamic constraint. We further apply the approach to a model of PKA-calcineurin signaling in synaptic long-term depression, demonstrating its applicability to a more complex biochemical reaction network.
\end{abstract}

\section{Introduction}
Subcellular signaling pathways regulating, e.g., cell division or synaptic plasticity, are often modeled as \textit{dynamical networks of reactions}. The inference of such networks is essential to understand, predict, and control the processes they describe. 

\begin{figure}[htbp]
    \centering
    \begin{subfigure}[c]{0.35\textwidth}
        \centering
        \includegraphics[width=\textwidth]{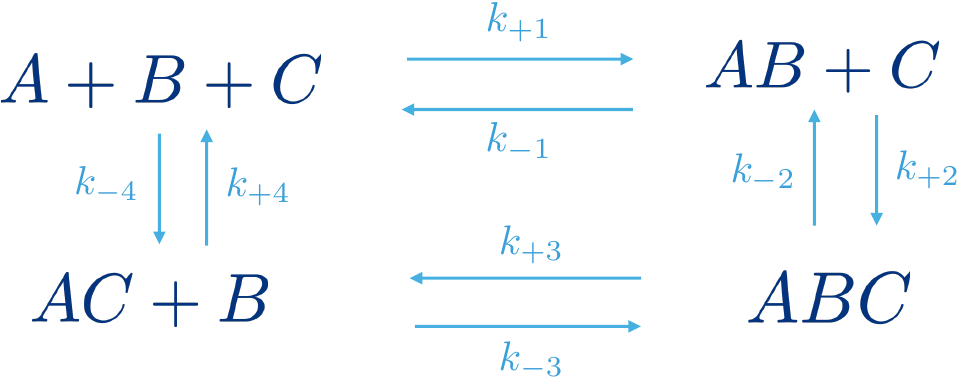}
        \caption{Reaction Network Diagram.}
        \label{fig:reaction_graph}
    \end{subfigure}
    \hfill
    \begin{subfigure}[c]{0.55\textwidth}
        \centering
        \includegraphics[width=\textwidth]{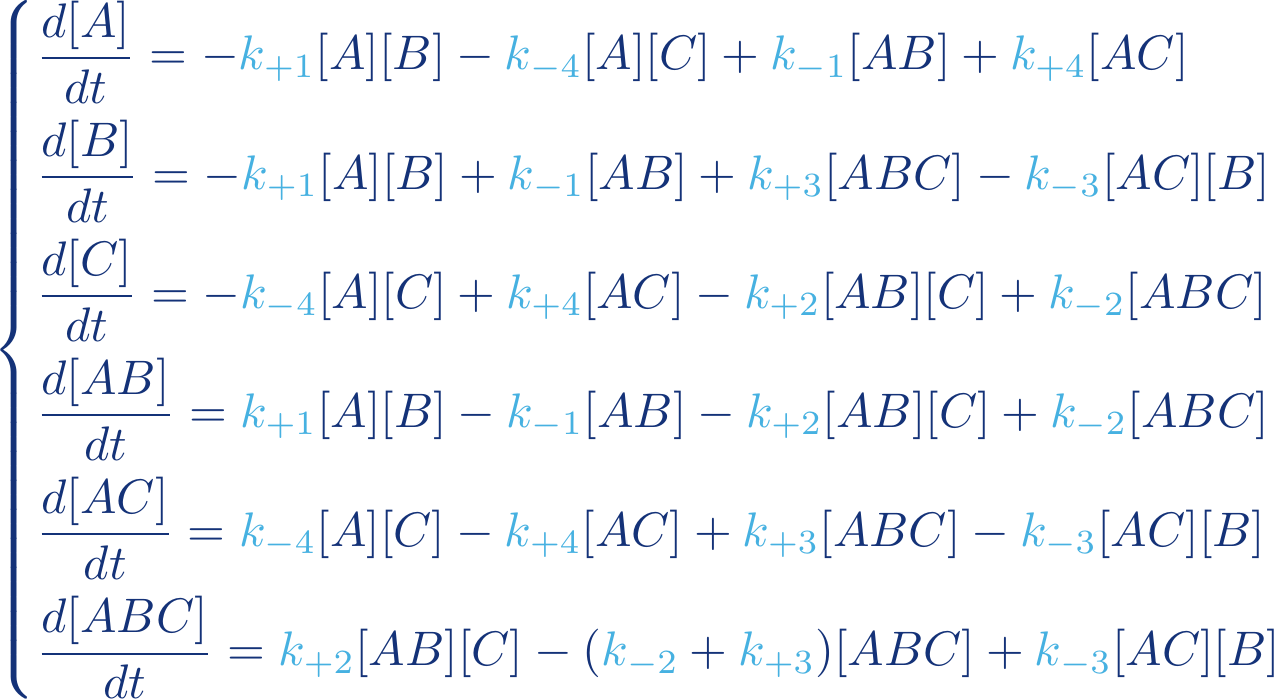}
        \caption{ODE model for the time evolution of the compounds' concentrations.}
        \label{fig:ode_model}
    \end{subfigure}
    \caption{One-cycle reversible reaction network and corresponding ODE model under mass-action kinetics. The chemical species are $A$, $B$, $C$, $AB$, $AC$, and $ABC$. The reaction rate constants are $k_{+j},k_{-j}$ for $j=1,2,3,4$. Note that in the reaction graph it is possible to ``walk'' from some states to other states following different reaction paths. For example, it is possible to ``walk'' from state $A+B+C$ to $ABC$ both through reaction $+1$ followed by reaction $+2$, or through reaction $-4$ followed by reaction $-3$.}
    \label{fig:graph_and_model}
\end{figure}

The models typically consist of a set of reacting proteins or protein complexes, where the reactions are represented by a graph, together with dynamical equations describing the rate of change of species concentration or activity (see Figure \ref{fig:graph_and_model} for an example). When constructing these models, it is important to take fundamental physical laws into account so that the system's behavior predicted by such models can actually occur in the real world, and not only in simulations. The dynamical equations are often based on the law of mass action together with the stoichiometric structure of the reaction network, which ensures conservation of mass in a closed system that is correctly specified. This gives rise to systems of ordinary differential equations. Another important principle is the law of detailed balance, which imposes thermodynamic constraints on the reaction rates in a closed system at thermodynamic equilibrium. The principle of detailed balance states that, at thermodynamic equilibrium each elementary process is equilibrated by its reverse process. Detailed balance thus demands that at thermodynamic equilibrium all net reaction fluxes vanish. This is important, in particular, when cycles appear in the system, or more in general, when there is more than a single reaction path connecting two nodes in the reaction network. Figure \ref{fig:reaction_graph} illustrates this type of structure. 

At thermodynamic equilibrium, regardless of the path taken between two nodes, the energy change must be the same. This imposes constraints on the admissible reaction rate constants. 
Wegscheider noted already in the early 1900 that, if the rate constants in a reaction network are chosen independently of each other, the system may reach a steady state that is not a thermodynamic equilibrium~\cite{wegscheider1902}. More specifically, although the concentrations of all chemical species remain constant in time, some reactions may still have nonzero net fluxes. Such behavior is incompatible with thermodynamic equilibrium in a closed system, as it would correspond to sustained net reaction fluxes at thermodynamic equilibrium.

Wegscheider found conditions (which are now referred to as the Wegscheider conditions of detailed balance) that, if fulfilled, are necessary and sufficient to guarantee that the equilibrium of the full network is compatible with the simultaneous equilibrium of each (isolated) elementary reaction \cite{gorban2015three}. For a reaction cycle, the Wegscheider conditions impose that the product of the reaction rate constants tranversing the cycle in one direction is the same as the product of the rate constants tranversing the cycle in the opposite direction. For example, for the cycle of four reversible reactions illustrated in Figure \ref{fig:reaction_graph}, this corresponds to
\begin{equation}
\label{eq:wegscheider_cycle}
k_{+1}\cdot k_{+2}\cdot k_{+3}\cdot k_{+4}=k_{-1}\cdot k_{-2}\cdot k_{-3}\cdot k_{-4}.
\end{equation}

Although this relationship is straightforward and could be, in principle, hard-coded into your model, problems arise when implementing it in larger, real-world, complex networks. This problem is particularly relevant when performing different types of parameter inference, such as during parameter estimation via optimization or Bayesian inference. One further complication arises when irreversible reactions are present in the model~\cite{gorban2011extended} and when we are dealing with an open system, since detailed balance apply to \textit{closed} systems of \textit{reversible} reactions at thermodynamic equilibrium. 

In this paper we suggest an approach to the problem of enforcing Wegscheider conditions that, to our knowledge, has not been explored before. Where others have focused on rewriting the model \cite{ederer2007thermodynamically}, describing relationships between parameters \cite{beard2004thermodynamic}, or incorporating the Wegscheider conditions as a hard constraint in the prior~\cite{jenkinson2010thermodynamically}, we add fictitious observations that enforce the detailed balance conditions. These fictitious observations are based on the fact that at thermodynamic equilibrium the net flux of all reversible reactions is zero. We first introduce this method for networks of reversible reactions. Later, we discuss how this can be extended to models with irreversible reactions and to open systems, which are not necessarily at thermodynamic equilibrium. In an optimization context, the fictitious observations are added to the loss function, promoting parameters that satisfy detailed balance, at least approximately. In the Bayesian framework, the fictitious observations modify the posterior density, promoting regions of the parameter space where the detailed-balance constraints are satisfied.

The remainder of the paper is organized as follows. In Section~\ref{sec:problem-statement}, we introduce the reaction network models considered in this work and the thermodynamic constraints imposed by detailed balance and the Wegscheider conditions. Section~\ref{sec:inference} then describes the maximum-likelihood and Bayesian approaches for inferring reaction rate constants from experimental data. Next, in Section~\ref{sec:inference}, we present the proposed approach based on fictitious observations, discuss its interpretation as a physics-informed prior, and describe its extension to models containing irreversible reactions and to open systems. Finally, in Section~\ref{sec:results}, we illustrate the proposed approach on a simple one-cycle reaction network and on a model of PKA-calcineurin signaling in synaptic long-term depression, before presenting our conclusions in Section~\ref{sec:conclusion}.

\section{Reaction network models and thermodynamic consistency}
\label{sec:problem-statement}

This section introduces the reaction network models considered in this work and formulates the corresponding parameter inference problem. We then describe the thermodynamic constraints imposed by detailed balance and the Wegscheider conditions.

\subsection{Reaction network models and parameter inference}
\label{sec:reaction-networks}

Consider a well-mixed biochemical system described by a reaction network with $n_x$ chemical species and $n_r$ reversible reaction pairs. For now, we assume that all reactions are reversible and refer to each forward-backward pair as a single reversible reaction. We denote the species in the system by $X_i$, with $i=1,\dots,n_x$. For each reversible reaction $j=1,\dots,n_r$, we denote by $R_j$ and $P_j$ the sets of indices of reactants and products, respectively. The stoichiometry of the $j$-th reaction is written as
\begin{equation}
\label{eq:reversible-reaction}
    \sum_{i\in R_j}|\nu_{R,ij}|X_i
    \mathrel{\mathop{\rightleftarrows}^{k_{+j}}_{k_{-j}}}
    \sum_{i\in P_j}|\nu_{P,ij}|X_i,
\end{equation}
where $|\nu_{R,ij}|$ and $|\nu_{P,ij}|$ are the stoichiometric coefficients of the reactants and products, respectively, and $k_{+j},k_{-j}>0$ are the forward and backward reaction rate (or kinetic) constants. We set $|\nu_{R,ij}|=0$ for $i\notin R_j$ and $|\nu_{P,ij}|=0$ for $i\notin P_j$.

We define the signed stoichiometric coefficient of species $X_i$ in reaction $j$ as
\begin{equation}
\label{eq:net-stoichiometric-coefficient}
    \nu_{ij}
    \coloneqq
    |\nu_{P,ij}|-|\nu_{R,ij}|.
\end{equation}
Thus, $\nu_{ij}<0$ if species $X_i$ is consumed by the forward reaction, $\nu_{ij}>0$ if it is produced, and $\nu_{ij}=0$ if its concentration is unchanged by the reaction. We collect these coefficients in the \textit{stoichiometric matrix}
\begin{equation}
\label{eq:stoichiometric-matrix}
    \mathbf{N}
    \coloneqq
    (\nu_{ij})_{
        \substack{i=1,\dots,n_x\\j=1,\dots,n_r}
    }
    \in\mathbb{R}^{n_x\times n_r}.
\end{equation}
The $j$-th column of $\mathbf{N}$ therefore describes the net change in the concentrations produced by one occurrence of the forward direction of reaction $j$.

In this work, we assume that the stoichiometry of the reaction network, hence its stoichiometric matrix $\mathbf
N$, is known. \textbf{The reaction rate constants} $k_{+j}$ and $k_{-j}$, $j=1,\dots,n_r$, \textbf{are} instead \textbf{unknown and our goal is to infer them from experimental data}. We usually perform parameter inference in terms of the logarithms of the reaction rate constants. We therefore define
\begin{align}
\label{def:theta}
     \theta_{+j}\coloneqq\log k_{+j},    \qquad \theta_{-j}\coloneqq\log k_{-j},
\end{align}

and collect the log-rate constants in the parameter vector
\begin{equation}
\label{eq:log-parameter-vector}
    \bft\coloneqq\left(        \theta_{+1},\dots,\theta_{+n_r},\theta_{-1},\dots,\theta_{-n_r}    \right)^\top\in\Theta\subseteq\mathbb{R}^{n_\theta},    \qquad    n_\theta=2n_r.
\end{equation}
Here, $\Theta$ denotes a set of admissible parameters. The original reaction rate constants can then be recovered as $k_{+j}=\exp(\theta_{+j})$ and $k_{-j}=\exp(\theta_{-j})$.

Let $x_i=[X_i]$ denote the concentration of species $X_i$, and define the concentration vector
\[
    \bfx\coloneqq(x_1,\dots,x_{n_x})^\top\in\mathbb{R}_{\ge0}^{n_x}.
\]
Under mass-action kinetics and given a fixed vector of log-rate constants $\bft$, the forward and backward reaction fluxes for
reaction $j=1,\dots,n_r$ are
\begin{equation}
\label{eq:f-b-fluxes}
    J_{+j}(\bfx;\bft)    =    k_{+j}\prod_{i=1}^{n_x}x_i^{|\nu_{R,ij}|}
    \qquad\text{and}\qquad J_{-j}(\bfx;\bft) = k_{-j}\prod_{i=1}^{n_x}x_i^{|\nu_{P,ij}|},
\end{equation}
respectively. The quantities $J_{+j}$ and $J_{-j}$ describe the rates at which reaction $j$ proceeds in the forward and backward directions, respectively. Their difference defines the net reaction flux
\[
    J_j(\bfx;\bft)\coloneqq J_{+j}(\bfx;\bft)-J_{-j}(\bfx;\bft).
\]
Thus, $J_j>0$ corresponds to a net flux in the forward direction, $J_j<0$ to a net flux in the backward direction, and $J_j=0$ means that the two directions balance exactly.  We collect the net reaction fluxes in the vector
\[
    \mathbf{J}(\bfx;\bft)\coloneqq
    \left(J_1(\bfx;\bft),\dots,J_{n_r}(\bfx;\bft)\right)^\top
    \in\mathbb{R}^{n_r}.
\]

In the absence of external inputs, the law of mass action gives the concentration dynamics
\begin{equation}
\label{eq:mass-action-ODE}
    \dot{\bfx}(t)=\mathbf{N}\,\mathbf{J}(\bfx(t);\bft).
\end{equation}
Equivalently, for each species $X_i$,
\[
\dot{x}_i(t) = \sum_{j=1}^{n_r}\nu_{ij}J_j(\bfx(t);\bft).
\]
Thus, each reaction contributes to the rate of change of $x_i$ through its net flux $J_j$, weighted by the corresponding stoichiometric coefficient $\nu_{ij}$.

Equation~\eqref{eq:mass-action-ODE} describes the reaction dynamics in the absence of external inputs or additional model terms. To accommodate experimental inputs and other contributions to the dynamics, we generalize the dynamical model as
\begin{equation}
\label{eq:ODEsystem}
\begin{cases}
    \dot{\bfx}(t)=\boldsymbol{f}(\bfx(t),\bfu,\bft),\\
    \bfx(t_0)=\bfx_0,
\end{cases}
\end{equation}
where $\bfu\in\mathbb{R}^{n_u}$ denotes a vector of inputs. When no additional inputs are present, this general model reduces to~\eqref{eq:mass-action-ODE}.

We use $\bfe$ as a compact notation for the experimental conditions, collecting experiment-specific quantities such as the initial condition $\bfx_0$ and the inputs $\bfu$. We therefore write $\bfx(t;\bft,\bfe)$ when we want to emphasize the dependence of the trajectory on both the model parameters and the experimental conditions. When the experimental conditions are fixed, we simply write $\bfx(t;\bft)$.

As an example, consider the reaction network shown in Figure~\ref{fig:graph_and_model}. For that system, the concentration vector is
\[\bfx=([A],[B],[C],[AB],[AC],[ABC])^\top\]
and the parameter vector of log-rate constants is
\begin{align*}
    \bft=\left(
        \log k_{+1},\log k_{+2},\log k_{+3},\log k_{+4},
        \log k_{-1},\log k_{-2},\log k_{-3},\log k_{-4}
    \right)^\top.
\end{align*}
Hence, $n_x=6$, $n_r=4$, and $n_\theta=8$.

Typically, we cannot observe the full state of the system $\bfx$. Rather, we measure a function $\bfg(\bfx)$ of the state, where $\bfg:\mathbb{R}^{n_x}\to\mathbb{R}^{n_y}$ is the output function. For example, a scalar output could be the concentration of a single species, $\bfg(\bfx)=x_1$, or the ratio of two concentrations, $\bfg(\bfx)=x_1/x_2$, whenever $x_2\neq0$.

In practice, these model outputs are observed only at discrete times and are affected by measurement error. Here, we consider experimental data consisting of noisy measurements collected at times $t_1,\dots,t_{n_t}$ and under experimental conditions $\bfe_1,\dots,\bfe_{n_e}$. We denote the observation collected at time $t_k$ under conditions $\bfe_\xi$ by $\bfy^{\rm obs}_{k,\xi}\in\mathbb{R}^{n_y}$. For a given parameter vector $\bft$, the corresponding model output is
\[
    \bfg\bigl(\bfx(t_k;\bft,\bfe_\xi)\bigr).
\]

Our goal is to infer the parameter vector $\bft$ from the collection of experimental data. To this end, we compare the model outputs $\bfg(\bfx(t_k;\bft,\bfe_\xi))$ with the observations $\bfy^{\rm obs}_{k,\xi}$ across all time points and experimental conditions, while accounting for a suitable measurement-error model.

At first sight, one might regard a parameter vector that reproduces the experimental observations well as a satisfactory solution to the inference problem. This is not necessarily the case. As we show in the numerical experiment in Section~\ref{sec:results-det-bal-violated}, parameters inferred from the data alone can reproduce the observed dynamics accurately while violating thermodynamic laws. The next subsection introduces thermodynamic constraints and explains why they must be taken into account. \textbf{The main goal of this work is thus to incorporate these constraints efficiently into the parameter inference procedure, avoiding parameter estimates that fit the experimental data but violate thermodynamic constraints.}

\subsection{Thermodynamic equilibrium and detailed balance}
\label{sec:detailed-balance}

As discussed in the previous subsection, agreement with the experimental data alone does not guarantee that the inferred reaction rate constants are physically admissible. We now introduce the thermodynamic constraints that reaction fluxes and rate constants must satisfy. 


For now, we restrict our attention to reversible reaction networks and consider experimental conditions $\bfe$ under which the solution of the ODE system~\eqref{eq:ODEsystem} converges to a strictly positive steady state
\begin{equation}
 \label{eq:steady-state}
    \bfx^{\rm eq}(\bft,\bfe)
    \coloneqq
    \lim_{t\to\infty}
    \bfx(t;\bft,\bfe)
    \in \mathbb{R}_{>0}^{n_x}.
\end{equation}
Importantly, convergence to a steady state does not, by itself, imply that the modeled system has reached thermodynamic equilibrium. For the closed reaction model~\eqref{eq:mass-action-ODE}, a steady state only requires
\[
\mathbf{N}\mathbf{J}(\bfx^{\rm eq};\bft)=\mathbf{0},
\]
which does not necessarily imply that the individual net reaction fluxes $J_j$ of the model vanish. Nonzero fluxes $J_j$ from different reactions may indeed balance each other so that the species concentrations $\bfx$ remain constant in time.

For a closed system of reversible elementary reactions, thermodynamic equilibrium imposes the stronger requirement that each forward reaction be exactly balanced by its corresponding backward reaction. Thermodynamic equilibrium imposes the requirement that each forward reaction be exactly balanced by its corresponding backward reaction. 
In particular, we say that a positive steady state $\bfx^{\rm eq}$ satisfies \textit{detailed balance} if
\begin{equation}
\label{eq:detailed-balance}
    J_{+j}(\bfx^{\rm eq};\bft)
    =
    J_{-j}(\bfx^{\rm eq};\bft),
    \qquad j=1,\dots,n_r.
\end{equation}
Equivalently, the net flux of each reversible reaction vanishes:
\begin{equation}
\label{eq:detailed-balance-vector}
    \mathbf{J}(\bfx^{\rm eq};\bft)=\mathbf{0}.
\end{equation}

Recalling~\eqref{eq:f-b-fluxes}, the detailed balance condition becomes
\begin{align}
\label{eq:equal-flux}
      k_{+j}\prod_{i=1}^{n_x}(x_i^{\rm eq})^{|\nu_{R,ij}|}
    =
    k_{-j}\prod_{i=1}^{n_x}(x_i^{\rm eq})^{|\nu_{P,ij}|},
    \qquad j=1,\dots,n_r.
\end{align}

Observe that the equalities in~\eqref{eq:equal-flux} involve both the equilibrium
concentrations and the reaction rate constants. Thus, the existence of a positive equilibrium satisfying these equalities imposes constraints on the rate constants. These
constraints are known as the \textit{Wegscheider conditions}
\cite{ederer2007thermodynamically,gorban2011extended}.

\subsection{Generalized Wegscheider conditions}
\label{sec:generalized-weg-cond}
To formulate Wegscheider conditions, recall that the columns of the stoichiometric matrix $\mathbf{N}$ describe the changes in the species concentrations caused by the individual reactions. By definition, a vector $\mathbf{c}=(c_1,\dots,c_{n_r})^\top$ in the null space of $\mathbf{N}$, denoted by $\ker(\mathbf{N})$, satisfies 
\[
\mathbf{N}\mathbf{c}=\mathbf{0}.
\]
Thus, $\mathbf{c}$ represents a combination of reaction directions whose overall stoichiometric effect is zero. When the entries of $\mathbf{c}$ are integers, this has the following interpretation: $c_j>0$ indicates that reaction $j$ is followed in its forward direction, $c_j<0$ that it is followed in its backward direction, and $|c_j|$ gives the corresponding multiplicity. After completing the combination of reactions, there is no net change in the amount of any species. Such a integer-valued combination defines a \textit{stoichiometric cycle}.

More generally, for every cycle $\mathbf{c}\in\ker(\mathbf{N})$, detailed balance requires
\begin{equation}
\label{eq:wegscheider-general}
    \prod_{j=1}^{n_r}\left(\frac{k_{+j}}{k_{-j}}\right)^{c_j}=1, \qquad \mathbf{c}\in\ker(\mathbf{N}).
\end{equation}
These are called the \textit{generalized Wegscheider conditions}. Note that it is sufficient to impose~\eqref{eq:wegscheider-general} for a basis of $\ker(\mathbf{N})$, since every other vector in $\ker(\mathbf{N})$ can be written as a combination of the basis vectors.

For a closed cycle in which each participating reaction appears once, condition~\eqref{eq:wegscheider-general} states that the product of the rate constants when going around the cycle in one direction must equal the product of the rate constants when going around the same cycle in the opposite direction. Intuitively, at thermodynamic equilibrium there cannot be a preferred direction around a closed reaction cycle. A preference for one direction would be incompatible with thermodynamic equilibrium.

For example, in the network shown in Figure~\ref{fig:reaction_graph}, the four reactions form a stoichiometric cycle, corresponding to $\mathbf{c}=(1,1,1,1)^\top\in\ker(\mathbf{N})$, with corresponding Wegscheider condition given in~\eqref{eq:wegscheider_cycle}.

Thus, although the rate constants describe individual reactions, they cannot in general be chosen independently: the rate constants associated with reactions belonging to the same stoichiometric cycle must satisfy the corresponding Wegscheider condition~\eqref{eq:wegscheider-general}.

\section{Inference of reaction rate constants}
\label{sec:inference}
Having introduced the reaction network model and the thermodynamic constraints, we now turn to the inference of the unknown reaction rate constants from experimental data. We first specify the observation model and corresponding likelihood, which provide the basis for both the maximum-likelihood and Bayesian approaches considered in this work.

\subsection{Observation model and likelihood}
\label{sec:observation-model-and-likelihood}

The experimental data used to infer reaction rate constants are typically assumed to correspond to noisy measurements of the model outputs. Before introducing the optimization-based and Bayesian approaches to parameter estimation in the following sections, we first specify the stochastic model used to describe the experimental observations.

We assume that each experimental observation $\bfy^{\rm obs}_{k,\xi}\in\mathbb{R}^{n_y}$  corresponds to a noisy measurement of the model output $\bfg(\bfx(t_k;\bft,\bfe_\xi))$. More precisely, we introduce the random measurement
\begin{equation}
\label{eq:observation-model}
    \mathbf{Y}_{k,\xi}    =
    \bfg(\bfx(t_k;\bft,\bfe_\xi))
    +
    \boldsymbol{\epsilon}_{k,\xi},
    \qquad
    k=1,\dots,n_t,\quad
    \xi=1,\dots,n_e,
\end{equation}
where $\boldsymbol{\epsilon}_{k,\xi}\in\mathbb{R}^{n_y}$ denotes the measurement error. We assume that the measurement errors are independent and identically distributed as
\begin{equation}
\label{eq:measurement-error}
    \boldsymbol{\epsilon}_{k,\xi}
    \stackrel{\mathrm{iid}}{\sim}
    \mathcal{N}\!\left(\mathbf{0},\sigma^2 I_{n_y}\right),
\end{equation}
where the noise standard deviation $\sigma>0$ is assumed to be known and fixed. Thus, conditional on the parameter vector $\bft$ and the specified experimental conditions, the observations are independent and normally distributed around the corresponding model outputs. Although more general measurement error models could be considered, we restrict our attention to this setting throughout this work.

Note that the dynamical model~\eqref{eq:ODEsystem} alone is deterministic. Conditional on $\bft$ and the specified experimental conditions $\bfe$, the stochasticity in the observation model arises entirely from measurement noise.

Under these assumptions, the conditional probability density of the full set of observations $\bfy^{\rm obs}$, given the parameter vector $\bft$, is
\begin{equation}
\label{eq:data_likelihood}
    \pi(\bfy^{\rm obs}\mid\bft)
    =
    (2\pi\sigma^2)^{-\frac{n_e n_t n_y}{2}}
    \exp\left\{
    -\frac{1}{2\sigma^2}
    \sum_{\xi=1}^{n_e}
    \sum_{k=1}^{n_t}
    \left\|
    \bfy^{\rm obs}_{k,\xi}
    -
    \bfg\bigl(\bfx(t_k;\bft,\bfe_\xi)\bigr)
    \right\|^2
    \right\},
\end{equation}
where $\|\cdot\|$ denotes the Euclidean norm in $\mathbb{R}^{n_y}$. When viewed as a function of $\bft$ for fixed observations $\bfy^{\rm obs}$, the density~\eqref{eq:data_likelihood} is the \emph{likelihood function}.

The likelihood $\bft\mapsto\pi(\bfy^{\rm obs}\mid\bft)$ quantifies how compatible different parameter values are with the observed data under the assumed observation model: parameter values for which $\pi(\bfy^{\rm obs}\mid\bft)$ is larger provide a better agreement between the model predictions $\bfg(\bfx(t;\bft,\bfe_\xi))$ and the observed data.

\subsection{Maximum likelihood estimation}
\label{sec:MLE}
A straightforward way to estimate the parameters of a model is to choose the parameter values $\bft$ that maximize the likelihood $\pi(\bfy^{\rm obs}\mid\bft)$. For a general treatment of maximum likelihood estimation and statistical inference, we refer to~\cite{casella2024statistical}. This optimization-based approach leads to the definition of the \textit{maximum likelihood estimator}
\begin{align*}
\widehat{\bft}_{\rm ML}
\in
\arg\max_{\bft\in\Theta}
\pi(\bfy^{\rm obs}\mid\bft).
\end{align*}
Equivalently, one can minimize the negative log-likelihood
\begin{align}
\label{eq:log-likelihood}
-\log\bigl(\pi(\bfy^{\rm obs}\mid\bft)\bigr)=\frac{n_e\cdot  n_t \cdot n_y}{2}\log(2\pi\sigma^2)+\frac{1}{2\sigma^2}\sum_{\xi=1}^{n_e}\sum_{k=1}^{n_t}\left\|\bfy^{\rm obs}_{k,\xi}-\bfg\left(\bfx(t_k;\bft,\bfe_\xi)\right)\right\|^2.
\end{align}
Thus, under the Gaussian noise model introduced in Section~\ref{sec:observation-model-and-likelihood}, minimizing~\eqref{eq:log-likelihood} is equivalent to minimizing the loss function 
\begin{align}
\label{eq:L-MLE}
    L_{\rm MLE}(\bft)\coloneqq \sum_{\xi=1}^{n_e}\sum_{k=1}^{n_t}\left\|\bfy^{\rm obs}_{k,\xi}-\bfg\left(\bfx(t_k;\bft,\bfe_\xi)\right)\right\|^2,
\end{align}
corresponding to the sum of squared residuals. Indeed, the first term in~\eqref{eq:log-likelihood} is constant with respect to $\bft$, while the second differs from~\eqref{eq:L-MLE} only by the positive constant multiplicative factor $1/(2\sigma^2)$.

This nonlinear least-squares problem can be addressed using algorithms such as  Gauss-Newton or Levenberg-Marquardt algorithms~\cite{gabor2015robust}, which exploit the derivatives of the residuals with respect to the parameters. For ODE models, these derivatives can be approximated by finite differences or computed by solving sensitivity equations~\cite{eriksson2026sensitivity}. 

\paragraph{Limitations of the optimization approach.}
The loss function $L_{\rm MLE}(\bft)$ may be nonconvex, and therefore local optimization algorithms may converge to different solutions depending on the initial guess. A multistart strategy, in which the optimization is repeated from several initial parameter vectors, can help explore different local minima, although it does not guarantee identification of a global minimum.

A further difficulty in optimization-based parameter estimation is \textit{nonidentifiability}, which arises when the available observations do not uniquely determine the parameter vector $\bft$. We distinguish between \textit{practical} and \textit{structural nonidentifiability}. 

Practical nonidentifiability arises when the available data are not sufficiently informative, for example because the measurements are sparse or affected by substantial measurement noise. In this case, different parameter values may lead to similar model predictions and therefore provide approximately equally good fits to the observed data.

Structural nonidentifiability, on the other hand, is an intrinsic property of the model and observation setup. It occurs when distinct parameter vectors lead to exactly the same model outputs under the considered experimental conditions, even in the idealized setting of noise-free observations. More precisely, there may exist a set $\mathcal{S}\subset\Theta$ such that
\[\bfg(\bfx(t;\bft,\bfe_\xi))=
\bfg(\bfx(t;\widetilde{\bft},\bfe_\xi))\]
for all $\bft,\widetilde{\bft}\in\mathcal{S}$, for all relevant times $t$ and experimental conditions $\bfe_\xi$. In such a case, the observations cannot distinguish between the parameter vectors in $\mathcal{S}$. Consequently, even with arbitrarily many noise-free measurements, it would not be possible to identify a unique parameter vector from the observed model outputs.

Nonidentifiability also affects the geometry of the optimization problem. If many parameter vectors provide the same or a very similar fit to the data, the loss function $L_{\rm MLE}(\bft)$ may be flat, or nearly flat, along certain directions in parameter space. As a consequence, optimization algorithms may converge slowly or return substantially different parameter estimates depending on the initialization and numerical tolerances. More importantly, in the presence of nonidentifiability, a single optimizer does not fully describe the range of parameter values that are compatible with the observed data.

A further limitation of using maximum likelihood estimation alone is that it provides a point estimate but does not, by itself, characterize the uncertainty associated with the inferred parameters. Additional techniques, such as profile likelihood or bootstrap methods, can be used for this purpose.

The Bayesian approach, introduced in the next section, provides a different framework for parameter inference. By considering the full \textit{posterior distribution} of the parameters, it allows us to characterize parameter uncertainty and to expose the effects of nonidentifiability on the inferred parameter values. The resulting parameter uncertainty can later also be propagated to predictions from the model.

\subsection{Bayesian parameter estimation and inverse uncertainty quantification}
\label{sec:bayes-param-estim}
In Bayesian inverse uncertainty quantification, we infer the parameter vector $\bft$ by combining prior information on $\bft$ with new information obtained from the observed data $\bfy^{\rm obs}$. The prior information is encoded in the \textit{prior distribution}, with density $\pi(\bft)$, assuming this density exists. The prior distribution incorporates information on the model parameter $\bft$ before observing the experimental data $\bfy^{\rm obs}$. For further details on Bayesian inference and statistical inverse problems, we refer to~\cite{gelman1995bayesian,kaipio2005statistical,casella2024statistical}.

For example, we might know from the literature on a specific reaction network model that its parameter $\bft$ should be approximately equal to a certain vector $\bft_{\rm pr}$. However, there might be large uncertainty in this estimate. We could therefore model the prior distribution, for instance, as a Gaussian distribution centered at $\bft_{\rm{pr}}$, with covariance matrix $\sigma_{\rm{pr}}^2I_{n_\theta}$, for a sufficiently large $\sigma_{\rm{pr}}$ to reflect this uncertainty: $\bft\sim\mathcal{N}(\bft_{\rm{pr}},\sigma_{\rm{pr}}^2I_{n_\theta})$. Thus,
\begin{equation*}
    \pi(\bft)\propto \exp\left\{-\frac{1}{2\sigma_{\rm{pr}}^2}\|\bft-\bft_{\rm{pr}}\|^2\right\}.
\end{equation*}
Here, $\sigma_{\rm pr}>0$ controls the spread of the prior. In particular, larger $\sigma_{\rm pr}$ corresponds to larger prior uncertainty in the parameter.

The Bayesian framework allows us to update the prior information on the parameter vector $\bft$ by incorporating the information from the observed data $\bfy^{\rm obs}$. This updated information is encoded in the \textit{posterior distribution}, which is the conditional distribution of $\bft$ given $\bfy^{\rm obs}$. By Bayes' theorem, its density is given by
\begin{align}
\label{eq:posterior-density}
        \pi(\bft\mid\bfy^{\rm obs})=\frac{\pi(\bft)\cdot \pi(\bfy^{\rm obs}\mid\bft)}{\displaystyle\int_{\Theta}\pi(\widetilde{\bft})\pi(\bfy^{\rm obs}\mid\widetilde{\bft})\,\mathrm{d}\widetilde{\bft}}
    \propto\pi(\bft)\cdot \pi(\bfy^{\rm obs}\mid\bft),
\end{align}
where the likelihood is given by~\eqref{eq:data_likelihood}. The integral in the denominator ensures that the posterior density integrates to one.

\paragraph{Maximum a posteriori estimation.}
Within the Bayesian framework, we can obtain a point estimate for the parameter $\bft$ by maximizing the posterior density. A maximizer of the posterior density~\eqref{eq:posterior-density} is called a \textit{maximum a posteriori} (MAP) estimate. This can equivalently be obtained by solving the optimization problem
\begin{align}
\label{eq:find-MAP}
    \min_{\bft\in\Theta}\left\{-\log\pi(\bfy^{\rm obs}\mid\bft)-\log\pi(\bft)\right\}.
\end{align}
Under the Gaussian observation model introduced in Section~\ref{sec:observation-model-and-likelihood}, this is equivalent to minimizing the loss function
\begin{align}
\label{eq:MAP-loss}
    L_{\rm MAP}(\bft)
    \coloneqq
    \frac{1}{2\sigma^2}L_{\rm MLE}(\bft)
    -\log\pi(\bft),
\end{align}
where $L_{\rm MLE}(\bft)$ is the sum of squared residuals loss given in~\eqref{eq:L-MLE}.

Note that the MAP approach still shares some of the difficulties encountered in maximum likelihood estimation. In particular, nonidentifiability may persist. Although the prior $\pi(\bft)$ can act as a regularizer and help reduce the ambiguity in the parameter values, the posterior distribution may remain nearly flat along nonidentifiable directions, especially when the prior is weakly informative. If an informative prior resolves such an ambiguity, the additional information comes from the prior rather than from the observed data. Moreover, the MAP estimate only provides a single representative point of the posterior distribution and therefore does not, by itself, characterize the uncertainty in $\bft$.

\paragraph{Uncertainty quantification via the Bayesian posterior.}
These limitations concern primarily the use of a single point estimate to summarize the parameter inference problem. By considering the full posterior distribution $\pi(\bft\mid\bfy^{\rm obs})$ instead, we retain information about all parameter values $\bft$ that are compatible with the observed data $\bfy^{\rm obs}$, given the model and the prior information $\pi(\bft)$. In particular, nonidentifiable or weakly identifiable directions can be represented by extended regions of high posterior probability, rather than forcing the inference toward a single parameter vector. At the same time, the spread and shape of the posterior distribution provide a direct characterization of parameter uncertainty. Thus, although Bayesian inference does not eliminate nonidentifiability or the associated computational challenges, it provides a natural framework for quantifying the parameter uncertainty for a given model that cannot be captured by a single optimization-based estimate. Uncertainty in the parameters can be propagated to model predictions, yielding posterior distributions and credible intervals for predicted quantities. These predictions can then be compared with experimental observations to falsify or validate the model.

\paragraph{Sampling from the posterior with Markov chain Monte Carlo methods.}

Evaluating the posterior density requires the computation of the integral in the denominator of~\ref{eq:posterior-density}. This can be computationally prohibitive, especially when the dimension $n_\theta$ of the parameter space is large, which is typically the case in applications. An alternative to a direct computation of the (normalized) posterior density is Monte Carlo sampling, which allows us to characterize the posterior distribution by generating \textit{random samples} from it.

We consider, in particular, Markov chain Monte Carlo (MCMC) methods, which generate a Markov chain of random parameter vectors $\bft^{(1)},\bft^{(2)},\bft^{(3)},\dots$ whose distribution converges, under mild assumptions, to the posterior $\pi(d\bft|\bfy^{\rm obs})$. For a comprehensive treatment of MCMC and Monte Carlo methods in general, we refer to~\cite{robert1999monte}. Many MCMC methods require the posterior density $\pi(\bft\mid\bfy^{\rm obs})$ only up to a normalizing constant. That is, it is sufficient to evaluate the product $\pi(\bft)\cdot\pi(\bfy^{\rm obs}|\bft)$ of the prior and the likelihood, without computing the expensive integral in the denominator of~\eqref{eq:posterior-density}. 

The resulting MCMC samples can then be used to approximate posterior quantities of interest and to quantify uncertainty in model predictions. Strong parameter dependencies and nonidentifiability can nevertheless
make sampling challenging.

\section{Methodology: incorporating thermodynamic constraints in the parameter estimation procedure}
\label{sec:methodology}
Our goal in this work is to enforce the thermodynamic constraints into our parameter estimation procedure in an efficient way. A straightforward approach to guarantee that these conditions are satisfied consists of directly imposing the Wegscheiden conditions. In Section~\ref{sec:limitation-direct-wegscheider}, we discuss the limitations of this strategy, while in Section~\ref{sec:proposed-approach}, we describe our proposed approach to the problem based on fictitious observations. Section~\ref{sec:alternative-interpretation} discusses an alternative interpretation of the proposed approach based on physics-informed priors.

\subsection{Limitations of directly imposing the Wegscheider conditions}
\label{sec:limitation-direct-wegscheider}
One way to impose detailed balance consists of determining the Wegscheider conditions~\eqref{eq:wegscheider-general} that relate the model parameters to each other and directly using them to reduce the model. For example, in the reaction network model in Figure~\ref{fig:reaction_graph}, the Wegscheider conditions yield the relation~\eqref{eq:wegscheider_cycle}. From this, we can express one of the eight reaction rate constants in terms of the others. For example, we can define
\begin{align*}
    k_{-4}\coloneqq \frac{k_{+1}\cdot k_{+2} \cdot k_{+3} \cdot k_{+4}}{k_{-1} \cdot k_{-2} \cdot k_{-3}},
\end{align*}
and exclude it from the parameter estimation procedure. That is, we model the system using the reduced parameter vector
$$\bft=(\log k_{+1},\log k_{+2},\log k_{+3},\log k_{+4},\log k_{-1},\log k_{-2},\log k_{-3})^\top\in\mathbb{R}^7$$
instead of 
$$\bft=(\log k_{+1},\log k_{+2},\log k_{+3},\log k_{+4},\log k_{-1},\log k_{-2},\log k_{-3},\log k_{-4})^T\in\mathbb{R}^8.$$
In so doing, we not only reduce the dimension of the parameter space, and hence the computational complexity of the parameter estimation procedure, but we also guarantee that the Wegscheider conditions are automatically satisfied. 

However, this approach requires a modeling choice: we need to decide which rate constant to eliminate when implementing the Wegscheider conditions. In our example, we chose to eliminate $k_{-4}$, but we might as well have eliminated any of the other seven parameters. Although equivalent parameterizations describe the same constrained model, these choices can affect numerical conditioning and the resulting optimization or sampling problem. In a Bayesian setting, additional care is also required when specifying priors under different parametrizations.

A further limitation of directly imposing the thermodynamic constraints is that real-world biochemical networks often are large. As discussed in Section~\ref{sec:generalized-weg-cond}, finding the Wegscheider conditions reduces to finding a set of independent stoichiometric cycles, represented by vectors in the null space ${\rm ker}(\mathbf{N})$ of the stoichiometric matrix. Although this computation is feasible using standard linear-algebra methods, explicitly constructing and implementing the corresponding Wegscheider conditions can become cumbersome as the size of the reaction network and the dimension of the null space ${\rm ker}(\mathbf{N})$ increase. As an alternative to directly imposing the Wegscheider conditions, we propose an approach that operates at the level of the individual reversible reactions and does not require explicitly enumerating the cycles of the entire network. 

\subsection{Proposed approach: fictitious observation of the zero net reaction fluxes}
\label{sec:proposed-approach}

Here we present our approach to enforcing the thermodynamic constraints, which does not require determining the Wegscheider conditions explicitly.

We recall that, for a parameter vector $\bft$ and experimental conditions $\bfe$ (i.e., initial condition $\bfx_0$ and inputs $\bfu$), we denote the solution of the ODE system~\eqref{eq:ODEsystem} at time $t$ by $\bfx(t;\bft,\bfe)$. Throughout this section, we assume that, for the parameter values and experimental conditions under consideration, the solution of the ODE system converges to a steady state:
\begin{align*}
    \bfx^{\rm eq}(\bft,\bfe)\coloneqq\lim_{t\to\infty}\bfx(t;\bft,\bfe).
\end{align*}
In the following, we write $\bfx^{\rm eq}$ for $ \bfx^{\rm eq}(\bft,\bfe)$.

As explained in Section~\ref{sec:detailed-balance}, detailed balance requires the net flux of each of the $n_r$ reversible reactions in the networks to vanish at equilibrium:
\begin{align*}
    \mathbf{J}(\bfx^{\rm eq};\bft,\bfe)=\boldsymbol{0}.
\end{align*}
Note that these conditions apply separately to each pair of forward and backward reactions. Thus, \textbf{they can be evaluated from the steady state solution of the ODE without explicitly determining the cycles of the reaction network}.

To enforce detailed balance, we adapt the augmented-observation approach used to incorporate prior information in statistical inverse problems, as described in~\cite[Section 4.4]{kaipio2005statistical}. We choose a fixed set of fictitious experimental conditions $\bfe^{\rm obs}$ such that the corresponding steady state has strictly positive components $x^{\rm eq}_i>0$, $i=1,\dots,n_x$, for all parameter vectors $\bft\in\Theta$ under consideration. We then introduce \textit{fictitious observations} $\bfy^{\rm fict}=\boldsymbol{0}\in\mathbb{R}^{n_r}$, associated with the stochastic measurement model
\begin{equation}
\label{eq:fictitious_observation}
    \boldsymbol{Y}^{\rm fict}=\mathbf{J}(\bfx^{\rm eq};\bft,\bfe^{\rm fict})+\boldsymbol{\epsilon}^{\rm fict},
    \qquad
   \boldsymbol{\epsilon}^{\rm fict}\sim\mathcal{N}(\mathbf{0},\sigma_{\rm fict}^2I_{n_r}).
\end{equation}
That is, we pretend to be measuring the net reaction fluxes at steady state with additive measurement error $\boldsymbol{\epsilon}^{\rm fict}$. Since detailed balance requires that the value of each net flux is exactly $0$, we assign the observed value $\bfy^{\rm fict}=\boldsymbol{0}$.

The fictitious measurement noise $\boldsymbol{\epsilon}^{\rm fict}$ is assumed to be independent of $\bft$ and of the (real) experimental measurement errors. Its standard deviation $\sigma_{\rm fict}>0$ controls how strongly deviations from detailed balance (zero net fluxes) are penalized, and is chosen as part of the modeling procedure. We emphasize that the stochastic model in~\eqref{eq:fictitious_observation} is artificial and does not represent an additional source of experimental noise. The fictitious observation is not an additional experimental measurement, but an artifice used to encode physical information into the likelihood.

Under the model in~\eqref{eq:fictitious_observation}, the likelihood associated with the fictitious observation is
\begin{align*}
    \pi(\bfy^{\rm fict}\mid\bft)=(2\pi\varepsilon^2)^{-\frac{n_r}{2}}\cdot \exp\left\{
        -\frac{\|\mathbf{J}(\bfx^{\rm eq};\bft,\bfe^{\rm fict})-\bfy^{\rm fict}\|^2}{2\varepsilon^2}
    \right\}.
\end{align*}
Since we always assign the value $\boldsymbol{0}$ to the fictitious observations, this becomes
\begin{align*}
    \pi(\bfy^{\rm fict}=\boldsymbol{0}\mid\bft)=(2\pi\varepsilon^2)^{-\frac{n_r}{2}}\cdot \exp\left\{
        -\frac{\|\mathbf{J}(\bfx^{\rm eq};\bft,\bfe^{\rm fict})\|^2}{2\varepsilon^2}
    \right\}.
\end{align*}

We refer to the observation model obtained by combining the experimental observations $\bfy^{\rm obs}$ with the fictitious observation $\bfy^{\rm fict}=\boldsymbol{0}$ as the \textit{augmented model}. By independence of the experimental and fictitious measurement errors, the corresponding augmented likelihood is given by
\begin{align*}
        \pi((\bfy^{\rm obs},\bfy^{\rm fict}=\boldsymbol{0})\mid\bft)=\pi(\bfy^{\rm obs}\mid\bft)\cdot\pi(\bfy^{\rm fict}=\boldsymbol{0}\mid\bft) \propto \pi(\bfy^{\rm obs}\mid\bft)\cdot \exp\left\{
        -\frac{\|\mathbf{J}(\bfx^{\rm eq};\bft,\bfe^{\rm fict})\|^2}{2\varepsilon^2}
    \right\},
\end{align*}
where $\pi(\bfy^{\rm obs}\mid\bft)$ is the data likelihood defined in~\eqref{eq:data_likelihood}. The corresponding posterior distribution is then given by
\begin{align*}
    \pi(\bft\mid (\bfy^{\rm obs}, \bfy^{\rm fict}=\boldsymbol0))=\frac{\pi(\bft)\cdot \pi(\bfy^{\rm obs}\mid\bft) \cdot \pi(\bfy^{\rm fict}=\boldsymbol{0}\mid\bft)}{\int_\Theta \pi(\tilde \bft)\cdot \pi(\bfy^{\rm obs}\mid\tilde \bft) \cdot \pi(\bfy^{\rm fict}=\boldsymbol{0}\mid\tilde \bft)d\tilde \bft}\propto \pi(\bft\mid \bfy^{\rm obs})\cdot \exp\left\{-\frac{\|\mathbf{J}(\bfx^{\rm eq};\bft,\bfe^{\rm fict})\|^2}{2\varepsilon^2}\right\}.
\end{align*}

The additional factor $\exp\left\{-\frac{\|\mathbf{J}(\bfx^{\rm eq};\bft,\bfe^{\rm fict})\|^2}{2\varepsilon^2}\right\}$ introduced by our proposed approach changes the posterior weight assigned to different parameter vectors $\bft$, penalizing those for which detailed balance is violated. In fact, if detailed balance is approximately satisfied, so that $\mathbf{J}(\bfx^{\rm eq};\bft,\bfe^{\rm fict})$ is small relative to $\sigma_{\rm fict}$, this exponential factor is close to one and the corresponding parameter vector $\bft$ receives little penalty. If, instead, the detailed balance condition is strongly violated, the net-flux norm $\|\mathbf{J}(\bfx^{\rm eq};\bft,\bfe^{\rm fict})\|$ is large relative to $\sigma_{\rm fict}$, and the exponential factor becomes close to zero. Consequently, the resulting posterior weight associated with $\bft$ is heavily reduced.

The same construction can be applied in an optimization setting. Let $L_{\rm MLE}(\bft)$ denote the loss function defined in~\eqref{eq:L-MLE}. Ignoring additive constants independent of $\bft$, the negative log-likelihood of the augmented observation model is
\begin{align}
\label{eq}
L_{\rm augm}(\bft)\coloneqq \frac{1}{2\sigma^2}L_{\rm MLE}(\bft)+\frac{1}{2\sigma_{\rm fict}^2}\|\mathbf{J}\bigl(\bfx^{\rm eq};\bft,\bfe^{\rm fict}\bigr)\|^2.
\end{align}
Maximum likelihood estimation under the augmented model therefore corresponds to minimizing $L_{\rm augm}(\bft)$. Similarly, the corresponding MAP estimator is obtained by solving
\begin{align*}
    \min_{\bft\in\Theta}
    \left\{
        L_{\rm MAP}(\bft)
        +
        \frac{1}{2\sigma_{\rm fict}^2}
        \|\mathbf{J}(\bfx^{\rm eq};\bft)\|^2
    \right\},
\end{align*}
where $L_{\rm MAP}(\bft)$ is defined in~\eqref{eq:MAP-loss}. Thus, in the optimization formulation, the fictitious observations appear as an additional penalty term that drives the net equilibrium fluxes toward zero.

The choice of $\sigma_{\rm fict}$ determines the strength of this penalization. A small value of $\sigma_{\rm fict}$ strongly penalizes nonzero equilibrium fluxes and therefore provides a close approximation to the hard constraint
\begin{align*}
    \mathbf{J}\bigl(\bfx^{\rm eq};\bft,\bfe^{\rm fict})=\boldsymbol{0}.
\end{align*}
For any fixed $\sigma_{\rm fict}>0$, however, the constraint remains soft rather than imposing exact equality. Conversely, a larger value of $\sigma_{\rm fict}$ allows larger deviations from detailed balance and results in a weaker constraint. Thus, $\sigma_{\rm fict}$ can be interpreted as a tolerance controlling how closely the detailed balance condition is required to hold. In practice, its value should also take into account the scale of the individual fluxes: if the different components of $\mathbf{J}$ have substantially different scales, the isotropic covariance $\sigma_{\rm fict}^2I_{n_r}$ can be replaced by a more general fixed positive definite covariance matrix in order to penalize the different fluxes appropriately.

\subsection{Interpretation of the proposed approach as physics-informed prior}
\label{sec:alternative-interpretation}

The approach presented above is formulated in terms of fictitious observations and therefore as an augmentation of the likelihood. However, the same additional factor can instead be incorporated into the prior distribution. More specifically, we can define a thermodynamically informed prior with density
\begin{align*}
\tilde\pi(\bft)\propto\pi(\bft)\cdot\exp\left\{-\frac{\|\mathbf{J}\bigl(\bfx^{\rm eq};\bft,\bfe^{\rm fict})\bigr)\|^2}{2\sigma_{\rm fict}^2}\right\}.
\end{align*}
Combining this prior with the original likelihood $\pi(\bfy^{\rm obs}\mid\bft)$ yields the same posterior distribution as the augmented model. In this interpretation, the additional factor encodes prior physical information through the forward model and its equilibrium fluxes, rather than specifying preferred parameter values directly. For this reason, we interpret $\tilde \pi(\bft)$ as a \textit{physics-informed prior}.

\subsection{Fictitious experiments in open models with irreversible reactions}
\label{sec:open-irreversible}
So far, we have worked under the assumption of a closed system containing only reversible reactions, such as the one showed in~\ref{fig:graph_and_model}. However, many real-world systems do not satisfy these assumptions: they may be open, e.g., have chemostatted species or external incoming fluxes, and they may also contain irreversible reactions. Nevertheless, the kinetic constant for an elementary reaction should, in principle\footnote{under the same physical conditions, such as temperature, pressure, and solvent.}, be independent of the reaction-network context in which the reaction occurs,
provided of course that other species do not directly influence this reaction. Therefore, if a subset of interconnected reversible elementary reactions satisfies the Wegscheider conditions in one reaction-network context, the same rate constants should satisfy the corresponding Wegscheider conditions when the same subset of reactions is embedded in another context.

We make use of this fact of conserved Wegscheider conditions when we consider more complex models, which may be open and contain irreversible reactions. We do this by constructing a fictitious \emph{experiment} where we remove all irreversible reactions, together with the reactions that make the system open. In this experiment we therefore only have reversible reactions in a closed system. The model should thereby follow detailed balance and we can construct fictitious observations of zero net reaction flux, as described in Section~\ref{sec:proposed-approach}. These fictitious observations in the fictitious experiments enforce detailed balance and enforce the kinetic constants to follow Wegscheider conditions. We can then add other real-world experiments that include irreversible reactions and reactions that make the system open. In these settings the model will not follow detailed balance, but our approach guarantees that the Wegscheider conditions are satisfied.

\begin{figure}
    \centering
    \includegraphics[width=1\linewidth]{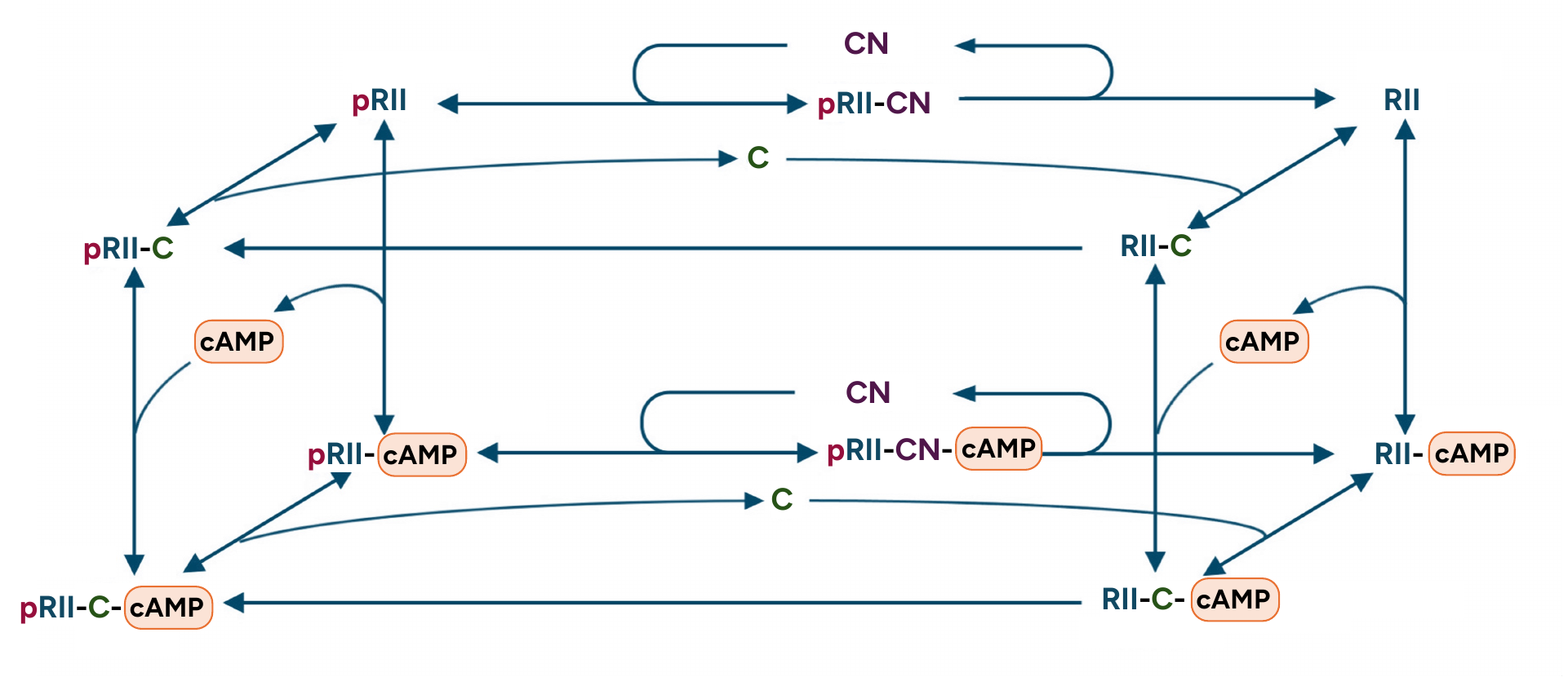}
    \caption{AKAP79 reaction network model for PKA–calcineurin signaling in synaptic long-term depression.
}
    \label{fig:AKAP79}
\end{figure}

In Section~\ref{sec:results-AKAP79}, we show this approach applied to the biochemical reaction network describing the PKA–calcineurin signaling in synaptic long-term depression~\cite{church2021akap79}. The network is represented in Figure~\ref{fig:AKAP79}, and is referred to as the AKAP79 model. Note that not all reactions in the system are reversible.

\section{Results}
\label{sec:results}

In this section, we present the results of parameter inference in the Bayesian framework described in Section~\ref{sec:bayes-param-estim} applied to the one-cycle model from Figure~\ref{fig:graph_and_model}. First, we show the results obtained when the thermodynamic constraints are not taken into account. In this case, we observe that, although the posterior samples provide a good fit to the experimental data, they violate the Wegscheider condition. We then repeat the Bayesian parameter inference using the proposed approach described in Section~\ref{sec:proposed-approach}. Finally, we apply the approach described in Section~\ref{sec:open-irreversible} for possibly open systems containing irreversible reactions applied to a model of PKA-calcineurin signaling in synaptic long-term depression.

\subsection{When detailed balance is ignored}
\label{sec:results-det-bal-violated}

We first perform parameter inference using the Bayesian posterior $\pi(\theta\mid\bfy^{\rm obs})$, which only takes into account the experimental data, without incorporating any thermodynamic constraints. 
We consider the one-cycle model from Figure~\ref{fig:graph_and_model} with one experiment consisting of synthetic time series measurements of the full state of the system (that is, $\bfg(\bfx)=\bfx$, hence $n_x=n_y$) at time points $t_k=k$, $k=1,2,\dots,9$. The synthetic experiment was generated by choosing upfront a parameter $\bft^{\rm true}$ corresponding (via~\eqref{def:theta}) to a set of rate constants $k_{\pm j}^{\rm true},j=1,\dots,4$, satisfying the Wegscheider condition~\eqref{eq:wegscheider_cycle} for the one-cycle model. Then, we computed the trajectory of the ODE system~\eqref{eq:ODEsystem} with parameter $\bft^{\rm true}$ and a fixed initial condition $\bfx_0$, and recorded the output $\bfg(\bfx(t))$ at time points $t_k=k$, for $k=1,\dots,9$. Later, we added independent gaussian noise $\boldsymbol{\epsilon}_k\sim\mathcal{N}(\boldsymbol{0},\sigma^2I_{n_x})$ to each output value $\bfg(\bfx(t_k))$, obtaining the (synthetic) observed data $\bfy^{\rm obs}=\bfg(\bfx(t_k))+\boldsymbol{\epsilon}_k$. Note that here we are only considering one experimental conditions (we generate only one time series), characterized by the initial condition $\bfx_0$, therefore we drop the index $\xi$ from~\eqref{eq:observation-model}.

\begin{figure}
    \centering
    \includegraphics[width=0.9\linewidth]{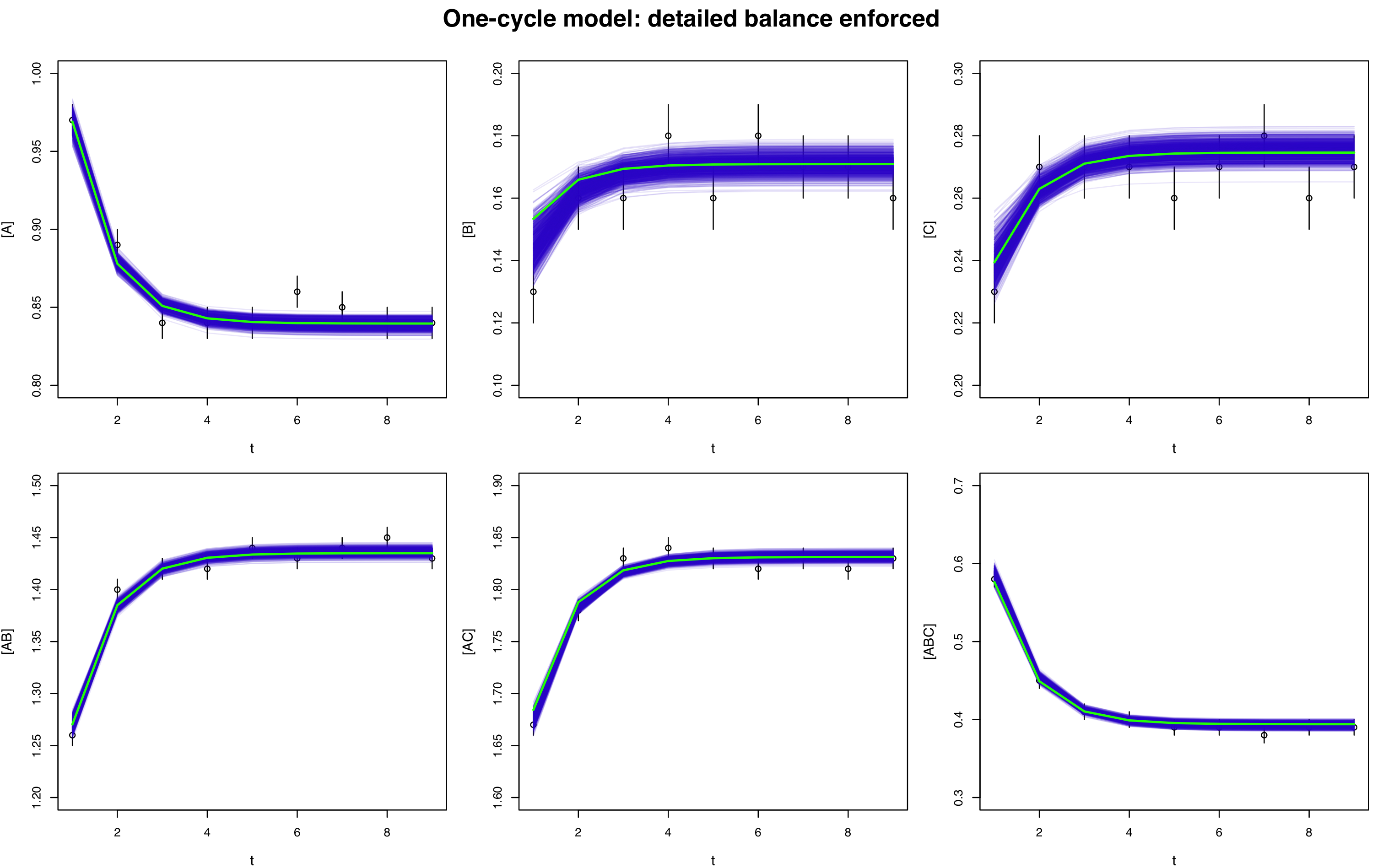}
    \caption{Experimental (synthetic) data and posterior trajectories for the one-cycle model without enforcing detailed balance. The experimental data are represented by black circles. The superimposed vertical segments show the corresponding standard errors. The posterior trajectories generated from the MCMC samples are plotted as blue curves (each blue curve corresponds to one MCMC sample $\bft^{(h)}$). The green curves represent the pointwise average of the posterior trajectories.}
    \label{fig:one-cycle_trajectories_det-bal-violated}
\end{figure}

We generate $1000$ samples $\{\bft^{(h)}\}$ from the posterior distribution $\pi(\theta\mid\bfy^{\rm obs})$ using the Random Walk Metropolis algorithm with an adaptive step size selection. The samples $\{\bft^{(h)}\}$ are then plugged into the ODE model~\eqref{eq:ODEsystem}, considering the same initial condition $\bfx_0$ as in the synthetic experiment from which the measurements were obtained. By simulating this model, we obtain trajectories for each of the six chemical species in the reaction network, which are showed in Figure~\ref{fig:one-cycle_trajectories_det-bal-violated}. 
From this figure, we observe that the simulated trajectories (blue curves) agree well with the experimental observations, and their pointwise average (green curves) closely follows the observed dynamics. Therefore, based only on the agreement between the model outputs and the data, the inferred parameter values \textit{appear} to provide a satisfactory description of the system.

\begin{figure}
    \centering
    \includegraphics[width=0.8\linewidth]{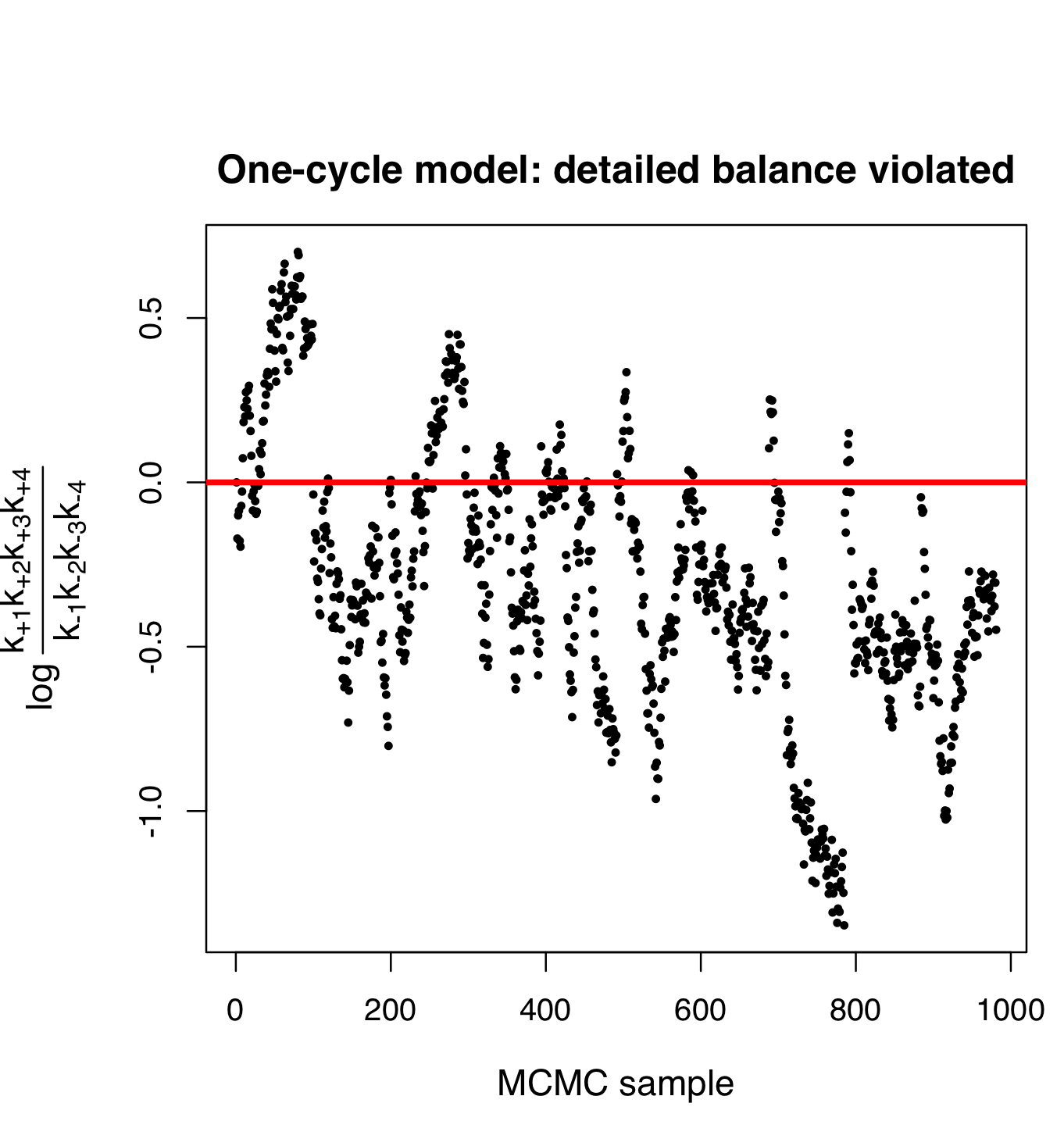}
    \caption{Log-Wegscheider residual defined in~\eqref{eq:log-wegscheider-residual} evaluated at the MCMC targeting the posterior distribution for the one-cycle model without enforcing detailed balance. Each point corresponds to one posterior sample $\bft^{(h)}$. The red horizontal line indicates the physically correct value $W(\bft)=0$.}
    \label{fig:one-cycle_weg-cond_det-bal-violated}
\end{figure}

However, when investigating the thermodynamic constraints, we observe that they are violated. This can be appreciated in Figure~\ref{fig:one-cycle_weg-cond_det-bal-violated}, where, for each of the MCMC samples $\bft^{(h)}$ targeting $\pi(\bft\mid\bfy^{\rm obs})$, we plot the value of the log-Wegscheider residuals, defined as 
\begin{equation}
\label{eq:log-wegscheider-residual}
    W(\bft)
    \coloneqq
    \log\left(\frac{k_{+1}k_{+2}k_{+3}k_{+4}}{k_{-1}k_{-2}k_{-3}k_{-4}}\right),
\end{equation}
where we recall that $\bft$ is related to $k_{\pm j}$ via~\eqref{def:theta}. As discussed in Section~\ref{sec:detailed-balance}, thermodynamic equilibrium requires that $W(\bft)=0$. However, the posterior samples obtained without enforcing detailed balance clearly violate this condition, and are therefore physically inadmissible.

\subsection{Enforcing detailed balance with the proposed approach}

We next perform parameter inference using the augmented model introduced in Section~\ref{sec:proposed-approach}. We recall that, by introducing the fictitious observations $\bfy^{\rm fict}$, we penalize parameter values $\bft$ for which the net reaction fluxes do not vanish at equilibrium.

\begin{figure}
    \centering
    \includegraphics[width=0.9\linewidth]{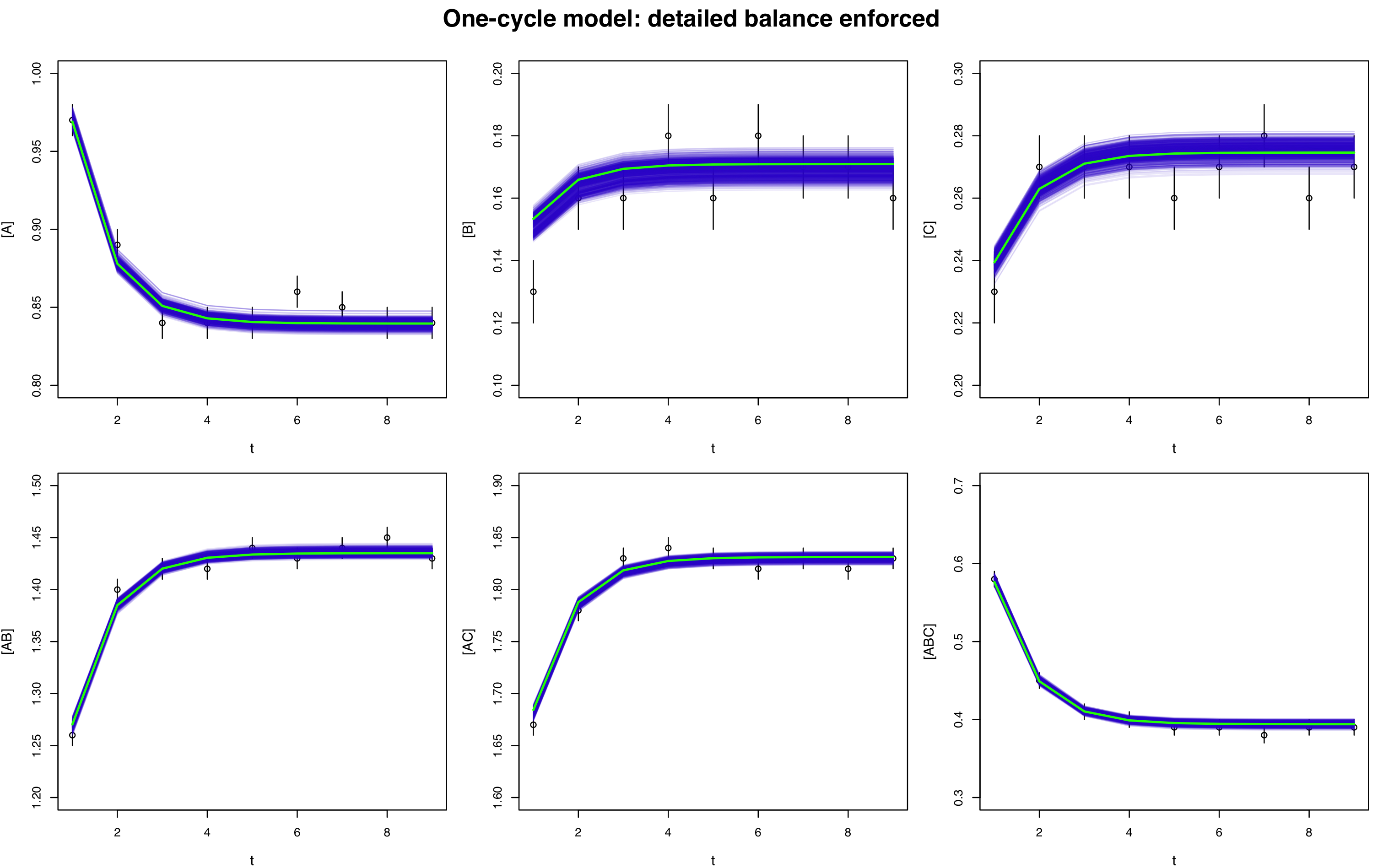}
    \caption{Experimental (synthetic) data and posterior trajectories for the one-cycle model obtained by enforcing detailed balance with the proposed approach described in Section~\ref{sec:proposed-approach}. The experimental data are represented with black circles. The superimposed vertical segments show the corresponding standard errors. The posterior trajectories generated from the MCMC samples are plotted as blue curves. The green curves represent the pointwise average of the posterior trajectories.}
    \label{fig:one-cycle_trajectories_det-bal-enforced}
\end{figure}

As shown in Figure~\ref{fig:one-cycle_trajectories_det-bal-enforced}, the model trajectories obtained from the augmented posterior continue to agree well with the experimental observations. In particular, incorporating the thermodynamic information does not substantially deteriorate the fit to the measured dynamics.

\begin{figure}
    \centering
    \includegraphics[width=0.8\linewidth]{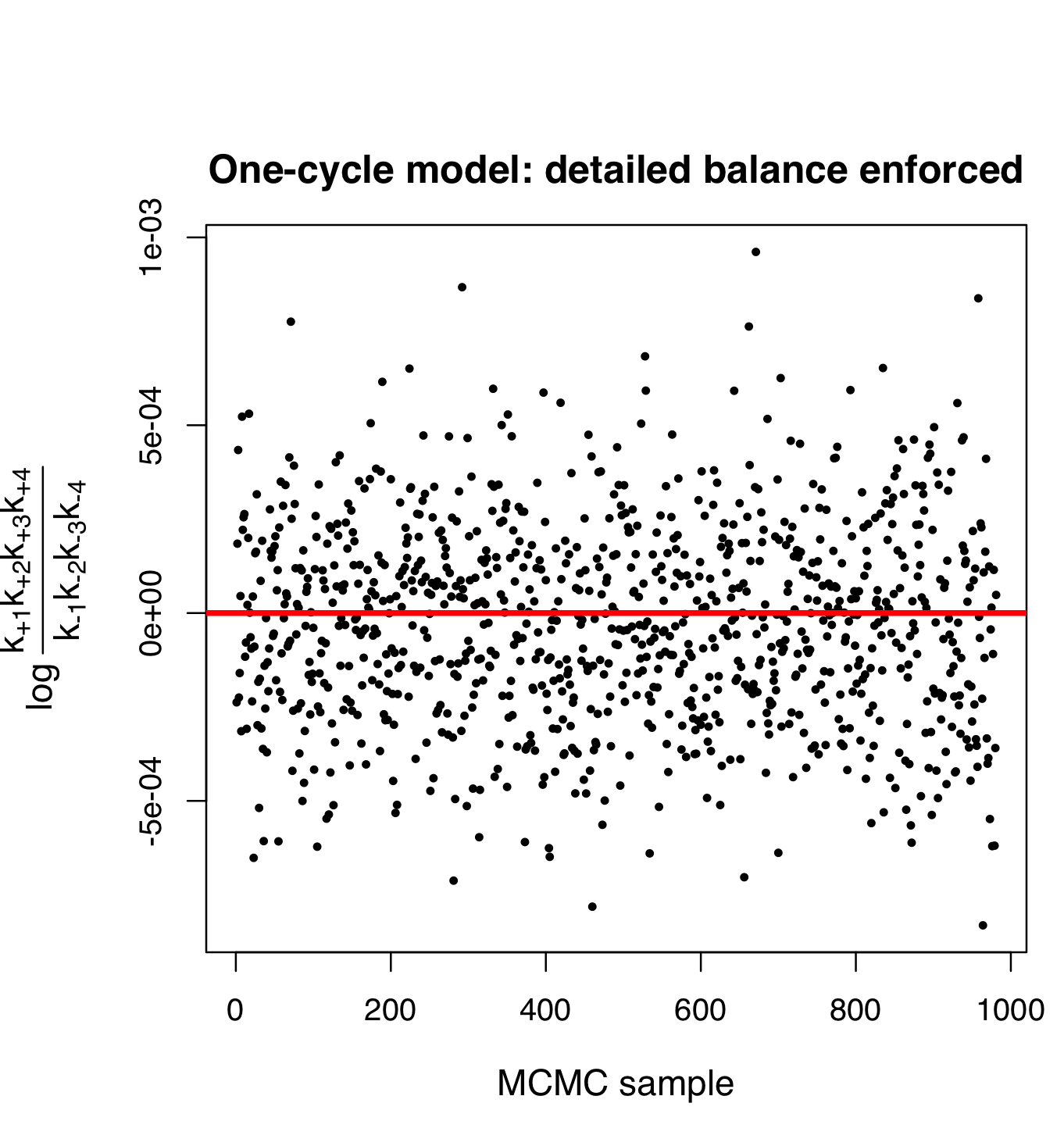}
    \caption{Log-Wegscheider residual defined in~\eqref{eq:log-wegscheider-residual} evaluated at the MCMC samples targeting the posterior distribution for the one-cycle model obtained enforcing detailed balance using the proposed approach described in Section~\ref{sec:proposed-approach}. Each point corresponds to one posterior sample $\bft^{(h)}$. The red horizontal line indicates the physically correct value $W(\bft)=0$.}
    \label{fig:one-cycle_weg-cond_det-bal-enforced}
\end{figure}

At the same time, the log-Wegscheider residuals $W(\bft^{(h)})$ shown in Figure~\ref{fig:one-cycle_weg-cond_det-bal-enforced} are concentrated close to the physically correct value $W(\bft)=0$. Thus, the posterior samples approximately satisfy the Wegscheider condition. Compared with the inference from Section~\ref{sec:results-det-bal-violated} where thermodynamic constraints are not taken into account, the proposed approach substantially reduces the thermodynamic inconsistency while preserving the agreement between the model predictions and the experimental data.

These results show that, for the one-cycle model considered here, the fictitious-observation approach successfully incorporates detailed balance into the parameter inference procedure. The method favors thermodynamically consistent parameter values without requiring an explicit reduction of the parameter space through directly imposing the Wegscheider condition.

\subsection{Enforcing detailed balance in an open network with nonreversible reactions}
\label{sec:results-AKAP79}
Finally, we apply the proposed approach to the AKAP79 model of PKA-calcineurin signaling in synaptic long-term depression~\cite{church2021akap79}. In contrast to the one-cycle example considered above, the AKAP79 model is open and contains irreversible reactions and therefore does not fall within the closed, fully reversible setting in which detailed balance can be imposed directly on the complete reaction network. Also here we construct synthetic experimental data, where we know that the Wegscheider conditions are satisfied. We do this for the full system including all reactions, i.e., also the irreversible, open reactions.

We apply the fictitious-experiment construction described in Section~\ref{sec:open-irreversible}. In particular, in addition to the synthetic experimental observations of the full system of reactions, we consider a fictitious closed system containing exclusively the reversible reactions of the AKAP79 model, and introduce fictitious observations requiring that their net fluxes vanish at equilibrium. These fictitious observations constrain the corresponding reaction rate constants to approximately satisfy the detailed balance condition, while the (synthetic)\footnote{Both the experimental data $\bfy^{\rm obs}$ and the fictitious observations $\bfy^{\rm fict}$ in these examples are generated synthetically, rather than derived from real experiments. For clarity, we emphasize that the experimental data $\bfy^{\rm obs}$ mimic real data that would be actually measured in a real-world experiment. Instead, the fictitious observations $\bfy^{\rm fict}$ are never measured: they only encode the information that the physically admissible value for the net reaction fluxes at equilibrium is zero.} experimental observations are described using the original AKAP79 model, including its irreversible reactions.

\begin{figure}
    \centering
    \includegraphics[width=0.9\linewidth]{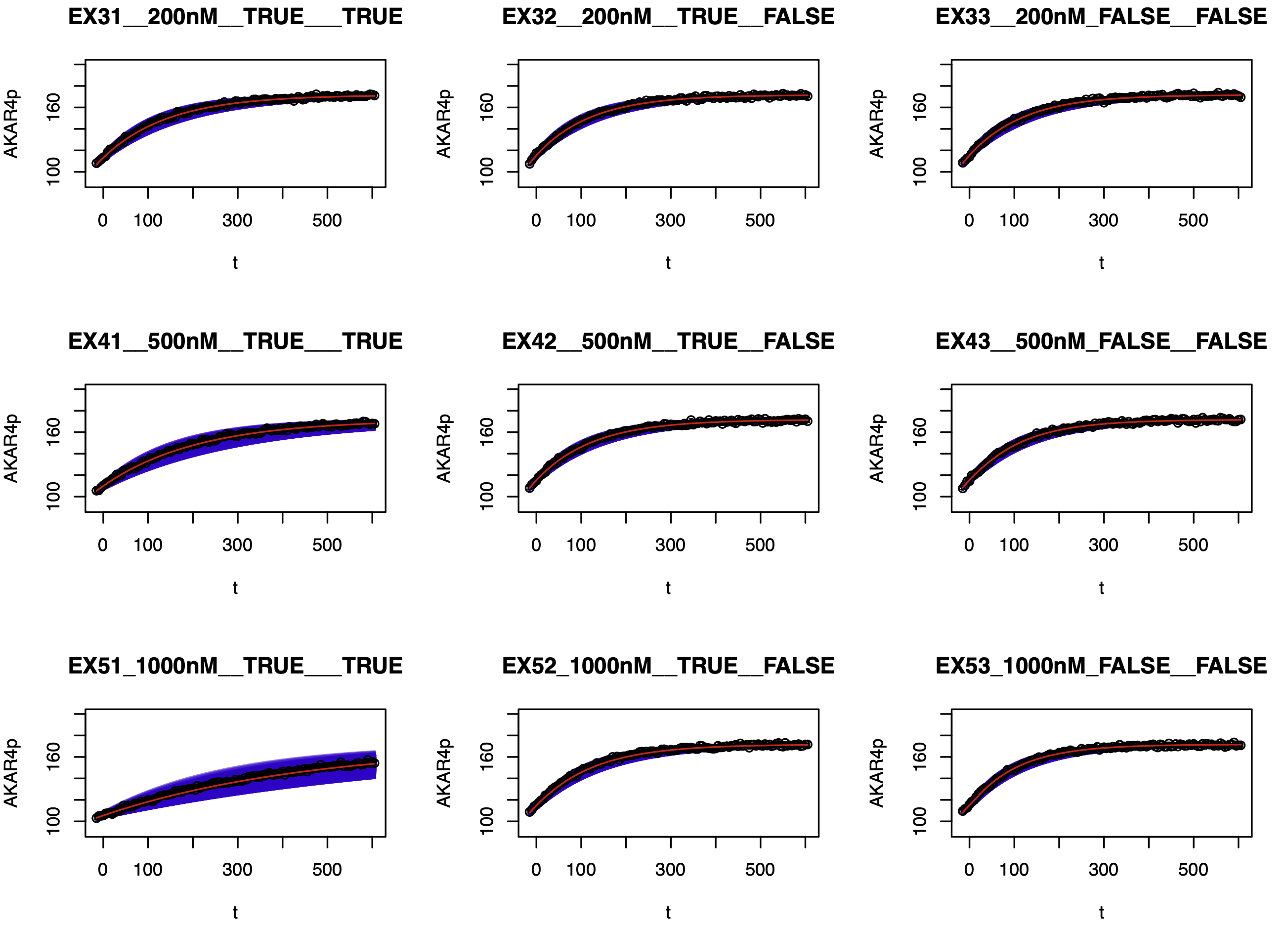}
    \caption{Experimental data and posterior trajectories for the AKAP79 model obtained by enforcing detailed balance. The experimental data are represented by black circles. The posterior trajectories generated from the MCMC samples are plotted as blue curves (each blue curve corresponds to one MCMC sample $\bft^{(h)}$). The red curves represent the pointwise average of the posterior trajectories.}
    \label{fig:traj-AKAP79}
\end{figure}

Figure~\ref{fig:traj-AKAP79} shows the posterior trajectories obtained from the augmented posterior together with the synthetic experimental observations. The inferred trajectories remain in good agreement with the observed dynamics. Thus, also for this more complex reaction network, incorporating the thermodynamic information through the fictitious experiment does not prevent the model from reproducing the experimental data.

\begin{figure}
    \centering
    \includegraphics[width=0.8\linewidth]{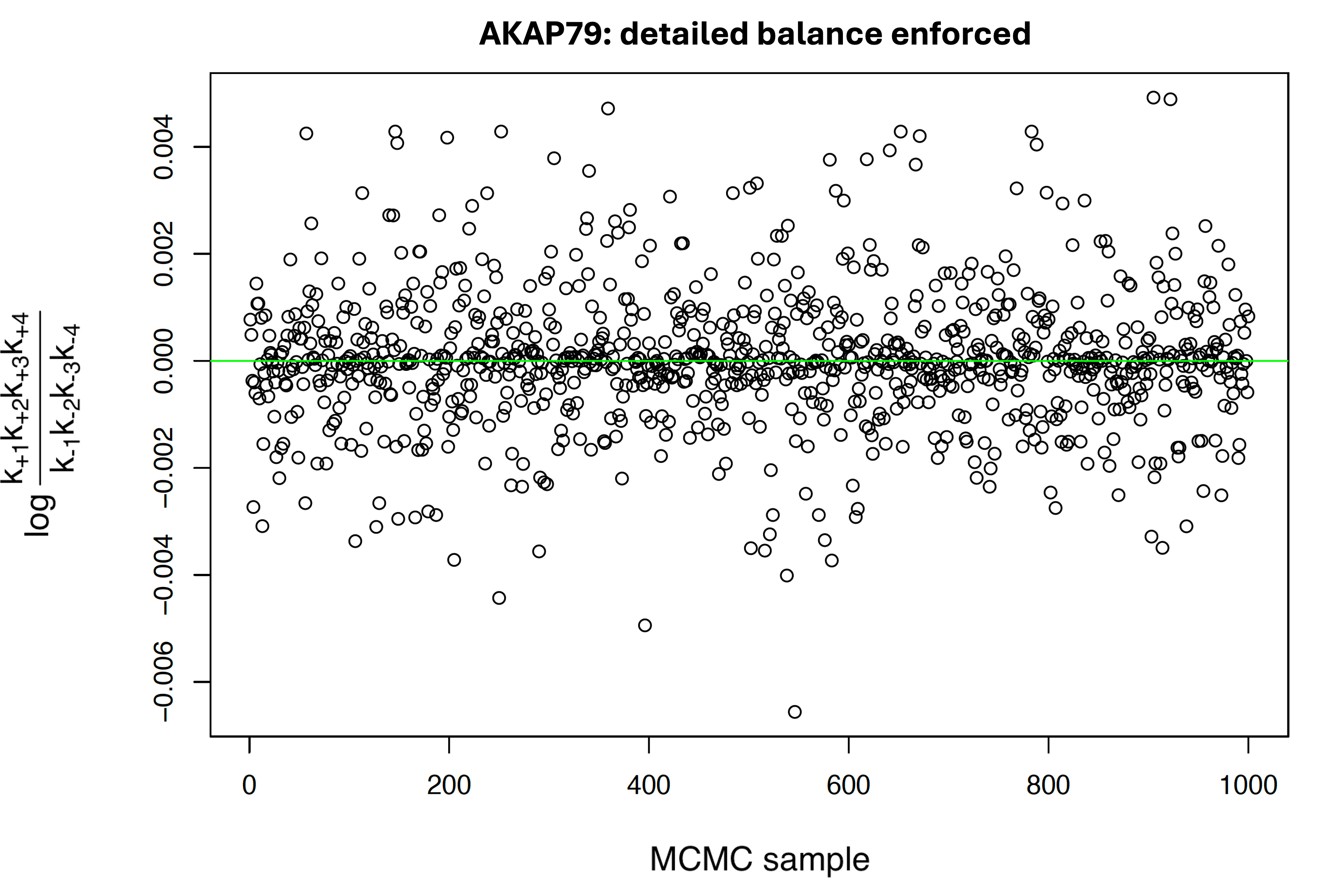}
    \caption{Log-Wegscheider residual defined in~\eqref{eq:log-wegscheider-residual} evaluated at the MCMC samples targeting the posterior distribution for the one-cycle model obtained enforcing detailed balance using the proposed approach described in Section~\ref{sec:proposed-approach}. Each point corresponds to one posterior sample $\bft^{(h)}$. The red horizontal line indicates the physically correct value $W(\bft)=0$.}
    \label{fig:weg-cond-AKAP79}
\end{figure}

To test the thermodynamic constraints, we evaluate the log-Wegscheider residuals associated with the reversible cycles that compose the AKAP79 model. In particular, these are the two cycles, one on the left and one on the right in Figure~\ref{fig:AKAP79}, which contain four reversible reactions each. The log-Wegscheider residuals $W(\bft)$ are thus of the same form as~\eqref{eq:log-wegscheider-residual}. Figure~\ref{fig:weg-cond-AKAP79} shows the values $W(\bft^{(h)})$ for each of the MCMC samples targeting the augmented posterior distribution where fictitious observations are included. We observe that these values are concentrated around the physically correct value $W(\bft)=0$, showing that the proposed approach effectively constrains the reaction rate constants toward thermodynamically consistent values, while still fitting the experimental data.

\section{Conclusion}
\label{sec:conclusion}

In this work, we considered the problem of incorporating thermodynamic constraints into the inference of reaction rate constants in biochemical reaction network models. A good agreement between model predictions and experimental observations does not, by itself, guarantee that the inferred reaction rate constants are physically admissible. In particular, for closed systems of reversible reactions, detailed balance imposes constraints on the reaction rate constants through the Wegscheider conditions.

Rather than explicitly deriving and imposing these conditions in parameter space, we proposed an approach based on augmenting the experimental data, including fictitious observations of zero net reaction fluxes at thermodynamic equilibrium. In the Bayesian setting, these observations introduce an additional factor in the posterior density that penalizes violations of detailed balance, while in an optimization setting they give rise to an additional penalty term in the loss function. The strength of the constraint is controlled by the standard deviation $\sigma_{\rm fict}$ of the fictitious observations. The same construction can alternatively be interpreted as introducing a physics-informed prior on the reaction rate constants.

On a simple reaction network containing a single cycle, we showed that posterior samples obtained using the experimental data alone can accurately reproduce the observed dynamics while substantially violating the Wegscheider condition. Incorporating the fictitious observations with the proposed approach preserves the agreement with the experimental data while yielding posterior samples that approximately satisfy the thermodynamic constraint. The approach therefore allows thermodynamic information to be incorporated without explicitly eliminating parameters or imposing the Wegscheider conditions through a reparameterization.

We further extended the approach to models that may contain irreversible reactions or describe open systems by introducing a fictitious closed experiment for the relevant reversible reactions. This allows the corresponding reaction rate constants to be constrained by the Wegscheider conditions without requiring the full model itself to satisfy detailed balance. The application to the AKAP79 model of PKA-calcineurin signaling in synaptic long-term depression illustrates the use of this construction in a more complex biochemical reaction network containing irreversible reactions.

The proposed approach provides a flexible way of incorporating thermodynamic information into parameter inference that does not require to identify the stoichiometric cycles in the reaction network. Its implementation requires the construction of suitable fictitious experimental conditions and the choice of the strength and scaling of the fictitious observation penalty. Further work could investigate systematic strategies for these choices and the performance of the approach in larger biochemical reaction networks.

\section*{Acknowledgments}
The project was funded from the Swedish e-Science Research Centre (SeRC). FM's research is supported by the Knut and Alice Wallenberg Foundation under grant KAW 2024.0327. The research of AK and OE were supported by the Swedish Foundation for Strategic Research (SSF) (grant number ITM24-0470) and Science for Life Laboratory. 

The computations and storage of large parameter-sample sets were enabled by resources provided by the National Academic Infrastructure for Supercomputing in Sweden (NAISS), partially funded by the Swedish Research Council through grant agreement no. 2022-06725.

FM gratefully thanks Prof. Daniela Calvetti and Prof. Erkki Somersalo for insightful discussions and valuable suggestions.

GPT-5.6 Sol (OpenAI) was used solely for grammatical and linguistic polishing. The authors are responsible for the scientific content of the manuscript.

\bibliographystyle{plain}
\bibliography{references}
\end{document}